\documentclass[amsmath,floatfix,twocolumn,superscriptaddress,citeautoscript]{revtex4-1}
\usepackage{subfigure}
\usepackage{siunitx}
\usepackage{amssymb}
\usepackage{amsmath}
\usepackage{graphicx}
\usepackage{array}
\usepackage{dcolumn}
\usepackage{psfrag}
\usepackage{bm}
\usepackage{color}
\usepackage{multirow}

\setcitestyle{super}

\begin{document}

\title{High-Throughput Surface Screening Reveals Energetically Competitive Low-Symmetry Facets in Wurtzite Semiconductors}

\author{Gustavo Madrigal}
\email{gustavo.adolfo.madrigal@upc.edu}
\affiliation{Departament de Física, Universitat Politècnica de Catalunya, 08034 Barcelona, Spain}
\affiliation{Barcelona Research Centre in Multiscale Science and Engineering, Universitat Politècnica de Catalunya, EEBE, 
Av. Eduard Maristany 10-14, Barcelona 08019, Spain}

\author{Cibr\'an L\'opez}
\affiliation{Departament de Física, Universitat Politècnica de Catalunya, 08034 Barcelona, Spain}
\affiliation{Barcelona Research Centre in Multiscale Science and Engineering, Universitat Politècnica de Catalunya, EEBE, 
Av. Eduard Maristany 10-14, Barcelona 08019, Spain}

\author{Jordi Llorca}
\affiliation{Barcelona Research Centre in Multiscale Science and Engineering, Universitat Politècnica de Catalunya, EEBE, 
Av. Eduard Maristany 10-14, Barcelona 08019, Spain}
\affiliation{Department of Chemical Engineering, Universitat Politècnica de Catalunya, EEBE, Av. Eduard Maristany 10-14, 
Barcelona 08019, Spain}

\author{Claudio Cazorla}
\email{claudio.cazorla@upc.edu}
\affiliation{Departament de Física, Universitat Politècnica de Catalunya, 08034 Barcelona, Spain}
\affiliation{Barcelona Research Centre in Multiscale Science and Engineering, Universitat Politècnica de Catalunya, EEBE, 
Av. Eduard Maristany 10-14, Barcelona 08019, Spain}
\affiliation{Institució Catalana de Recerca i Estudis Avançats (ICREA), Passeig Lluís Companys 23, 08010 Barcelona, Spain}

\begin{abstract}
Surfaces and interfaces govern the performance of wurtzite semiconductors in electronics, sensing, and solar-fuel conversion. 
To date, first-principles studies have focused mostly on high-symmetry low-index facets, owing to the computational cost and 
complexity of modelling low-symmetry high-index surfaces. In practice, however, surface formation energy is determined by the 
planar atomic density and the strength of the bonds severed upon cleavage, neither of which inherently excludes high-Miller-index 
surfaces from being energetically competitive. Here, we combine automated surface generation, machine-learning interatomic 
potential screening, and density functional theory to explore 226 clean, stoichiometric surface models of wurtzite GaN, LaN, 
and ZnO, spanning all Miller orientations with $|h|,|k|,|l| \le 3$ and their inequivalent terminations. We find that low-symmetry 
high-index prismatic facets are energetically competitive with respect to the conventional $m$- and $a$-planes: nine of the 
fifteen lowest-energy nonpolar surfaces are high-index, and the five best candidates of each compound lie within only 2.5--5.3\% 
of the minimum surface energy. Hybrid-functional calculations reveal, however, that this energetic proximity does not translate 
into similar electronic alignment. In GaN, the two most stable surfaces differ by just 0.08~J~m$^{-2}$ in formation energy but 
by almost 0.8~eV in vacuum-referenced band-edge positions, whereas in LaN the nearly degenerate $m$- and $a$-planes are 
separated by 0.16~eV. Wurtzite facets with similar formation energies may therefore exhibit very different ionization potentials 
and electron affinities. Our results enable the joint, systematic assessment of stability and band alignment across the facets 
that can actually occur in wurtzites, and provide a map of candidate surfaces for the rational design of related nanostructures 
and heterointerfaces.\\

\end{abstract}

\maketitle

\section{Introduction}
\label{sec:intro}
\begin{figure*}[t]
\centering
\includegraphics[width=\textwidth]{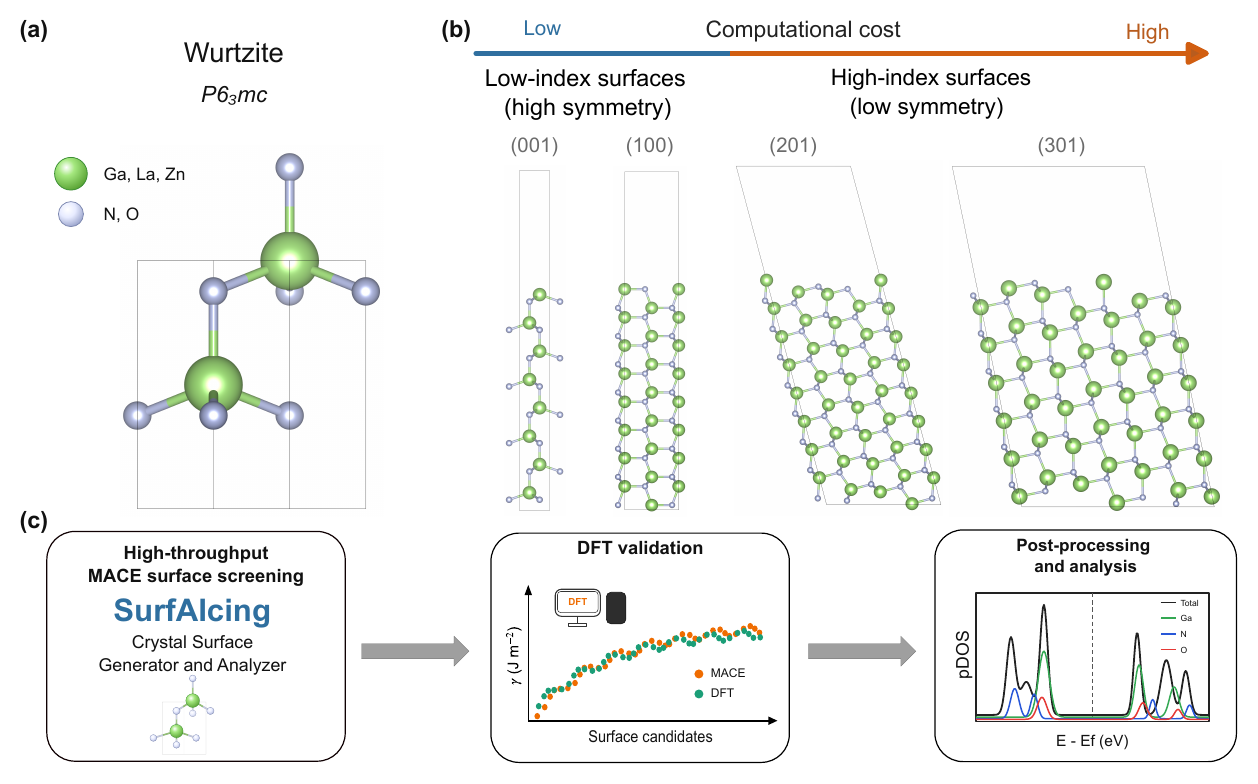}
\caption{(a)~Ball-stick representation of the wurtzite crystal structure with space group $P6_{3}mc$. 
         (b)~Representative low-index and high-index surface models, illustrating the increasing structural complexity and computational 
         demands associated with broader orientation sampling. 
         (c)~Computational workflow: surface generation with the \texttt{SurfAIcing} software \cite{surfaicing,Lopez2026}, MLIP-based 
         (MACE \cite{batatia2022mace,batatia2025foundation}) screening and structural relaxation, DFT refinement, and subsequent analysis 
         of surface energies and electronic properties.}
\label{fig1}
\end{figure*}

Wurtzite semiconductors form one of the most technologically versatile families of materials in applied physics. Group-III nitrides 
such as GaN and AlN underpin solid-state lighting \cite{Nakamura1994}, as well as deep-ultraviolet emitters \cite{Kneissl2019}, 
high-frequency radio-frequency amplifiers \cite{Mishra2008}, and power electronics \cite{Amano2018}. Wide-gap oxides such as ZnO are 
used in ultraviolet optoelectronics, transparent electronics, and sensing \cite{ozgur2005comprehensive}, and ZnO nanostructures have 
enabled piezoelectric nanogenerators for energy harvesting \cite{Wang2006}. What enables these diverse applications is the noncentrosymmetric 
hexagonal $P6_3mc$ structure (Fig.~\ref{fig1}a), whose tetrahedral coordination produces spontaneous polarization and a strong 
piezoelectric response \cite{Bernardini1997,dreyer2014absolute,rowberg2021structural}. These functional features give rise to 
two-dimensional electron gases at nitride heterointerfaces \cite{Ambacher1999}, enhance the performance of AlN-based piezoelectric 
devices \cite{Akiyama2009}, and trigger ferroelectricity in AlScN \cite{fichtner2019alscn}. Wurtzite-based materials are also relevant 
to solar fuel production; for example, GaN--ZnO solid solutions loaded with RuO$_2$ nanoparticles drive overall water splitting under 
visible light \cite{Maeda2005}.

In most of these applications, the functional behavior of wurtzite materials is controlled not by the bulk alone but by their surfaces 
and interfaces, where the crystal terminates and the coordination, bonding, and electrostatic environment of the atoms change abruptly. 
In electronic devices, contacts and heterojunctions govern carrier injection and transport: surface states can pin the Fermi level and 
set Schottky barrier heights \cite{Monch2001,VandeWalle2007}, and the polarization discontinuities at nitride heterointerfaces give rise 
to the two-dimensional electron gases exploited in high-frequency transistors \cite{Ambacher1999}. In ferroelectric materials, interfacial
bonding and electrostatic screening determine whether a polarization state is stable \cite{stengel2009enhancement}. In nanostructured 
wurtzite materials used for sensing, energy harvesting, and photocatalysis, the large surface-to-volume ratio makes the exposed facets 
the main sites of molecular adsorption, charge transfer, and chemical reaction \cite{Wang2004,Wang2006}, and these surfaces may exhibit 
markedly distinct behaviours \cite{debroye2017facet}. 

Two fundamental properties characterise surfaces. The first is the surface energy, which determines which orientations and terminations 
can be exposed under equilibrium and growth conditions \cite{dreyer2014absolute,li2015computing}. The second is the surface electronic 
structure, that is, the surface states and the electrostatic potential step between the material interior and the vacuum; this trait 
determines the ionization potential and electron affinity and hence the vacuum-referenced band-edge positions that control charge transfer 
at surfaces and interfaces \cite{hinuma2019band}. Knowing both properties across the facets that can actually occur is essential to 
understand crystal growth, morphology, and interfacial behaviour.

First-principles methods, and density functional theory (DFT) in particular \cite{Hohenberg1964,Kohn1965}, are well suited to quantify
both defining properties of a surface within a single atomistic framework. The surface energy follows from total-energy differences
between a relaxed slab and the bulk crystal, and it can be resolved by orientation and termination, extended to different chemical
environments through \textit{ab initio} thermodynamics \cite{Reuter2001}, and used to rank competing facets and construct equilibrium 
crystal shapes \cite{meyer2003density,dreyer2014absolute,li2015computing}. Structural relaxation, which accommodates the undercoordination 
of surface atoms, is obtained directly as part of those calculations. The electronic structure of the same relaxed slabs provides the 
surface density of states and the character and localization of surface states. Furthermore, the electrostatic potential step at the 
surface can be referenced to the vacuum level \cite{Kahn2016}, and combined with a bulk-to-surface potential lineup it yields the 
ionization potential, electron affinity, and vacuum-referenced band-edge positions \cite{VandeWalle1987,hinuma2019band}. 

Accuracy in these computed quantities depends on the choice of functional. Semilocal approximations tend to underestimate band gaps 
and misplace localized states because of self-interaction and delocalization errors \cite{mori2008localization}. Screened hybrid 
functionals \cite{Heyd2003} (e.g., the HSEsol variant \cite{Schimka2011,Krukau2006} designed for solids), on the other hand, provide 
a more reliable description of the electronic structure while remaining consistent with the accurate structural and energetic description 
of solid-optimized semilocal functionals. Such calculations have clarified the atomic structure, stability, and electronic properties 
of conventional wurtzite surfaces \cite{Staemmler2003PolarZnO,Du2008PolarZnO,Goniakowski2008}.
 
These simulation capabilities, however, have been exploited mostly for high-symmetry low-index surfaces, and the reasons are largely 
technical (Fig.~\ref{fig1}b). High-index orientations require large surface repeat cells that contain many atoms, and their low symmetry 
produces numerous inequivalent cleavage planes, so that each orientation gives rise to many candidate terminations instead of one or two. 
The number of structures to be generated, relaxed, and compared therefore grows rapidly with the range of Miller indices explored. Each 
of them is also demanding to converge: surface energies are sensitive to slab thickness and to the treatment of the two exposed faces 
\cite{Boettger1994,Fiorentini1996}. These difficulties are compounded when hybrid-functional calculations, whose cost is far
higher than that of semilocal ones, are required to characterize the electronic properties. Automated frameworks for slab generation 
and high-throughput surface calculations have been developed \cite{Sun2013,Ong2013,Tran2016}, but the systematic first-principles 
assessment of low-symmetry high-index surfaces, including their electronic properties, remains largely unexplored for wurtzite 
semiconductors. 

Addressing this knowledge gap is important because low-symmetry high-index facets are experimentally relevant: they have been reported 
to stabilize in ZnO nanopyramids with improved photocatalytic performance \cite{lim2021highindex}, and ZnO crystals show facet-dependent 
photoreduction \cite{debroye2017facet}. Without a systematic theoretical description, the stability and electronic alignment of these facets 
cannot be anticipated, and this lack of understanding may be hindering the rational design of wurtzite surfaces, nanostructures and
heterointerfaces for energy, information, and sensing applications.

In this study, we present a systematic first-principles exploration of the surface energetics and electronic properties of wurtzite 
GaN, LaN, and ZnO that extends beyond conventional high-symmetry low-index facets. Using the automated workflow of Fig.~\ref{fig1}c, 
the \verb!SurfAIcing! software \cite{surfaicing,Lopez2026} first generates clean, stoichiometric surface models for all Miller orientations 
with $|h|,|k|,|l| \leq 3$, together with their inequivalent terminations, yielding 226 candidates. A machine-learned interatomic
potential (MACE \cite{batatia2022mace,batatia2025foundation}) then provides preliminary energetic assessments and structural 
pre-relaxation, and semilocal DFT calculations on a selected subset of 97 candidates establish the final surface-energy rankings. 
The electronic structure of the five lowest-energy surfaces of each compound is then characterised with the HSEsol hybrid functional 
\cite{Schimka2011,Krukau2006}, including layer-resolved densities of states and vacuum-referenced band edges obtained via a bulk-to-surface 
potential lineup \cite{hinuma2019band}. 

We find that the lowest-energy nonpolar surfaces of each compound are all prismatic $(hk0)$ cuts, and that nine of the fifteen surfaces 
examined electronically are low-symmetry high-index, with energies within 2.5--5.3\% of the minimum of each compound. This energetic
proximity, however, does not translate into similar electronic alignment. In GaN, the two energetically most favourable surfaces
differ by only 0.081~J~m$^{-2}$ in formation energy but by almost $0.8$~eV in vacuum-referenced band-edge positions; in LaN, the two 
lowest energy, and nearly degenerate, surfaces differ by about $0.2$~eV, whereas the selected ZnO surfaces remain closely grouped. 
Formation energy alone, therefore, is not sufficient to globally assess the ionization potential and electron affinity of naturally 
occurring wurtzite surfaces. Our results show the value of extending the exploration of wurtzite surfaces beyond conventional facets 
and of assessing energetics and electronic properties jointly; they also provide a set of candidate surfaces for subsequent studies of 
reconstruction, adsorption, crystal morphology, and interface formation in energy, information, and sensing applications.

\section{Computational overview}
\label{sec:methods}
Our computational workflow combines automated surface generation and machine-learned interatomic potential (MLIP) screening with 
first-principles refinement and electronic-structure analysis (Fig.~\ref{fig1}c). The \verb!SurfAIcing! software \cite{surfaicing,Lopez2026} 
first generates surface structures and determines their equilibrium geometries with MACE \cite{batatia2022mace,batatia2025foundation}. 
Subsequently, semilocal DFT calculations establish the final relaxed structures and surface formation-energy rankings, while screened 
hybrid functionals characterise their electronic properties. Next, we provide a brief description of the technical details in our 
calculations.

\subsection{High-throughput surface generation and screening}
\label{sec:surface_generation}
Stoichiometric surface models of wurtzite GaN, LaN, and ZnO were generated from their corresponding bulk structures with space group 
$P6_3mc$. Miller orientations satisfying $|h|,|k|,|l|\leq 3$ were sampled. Throughout this work, orientations are expressed using the 
three-index notation $(hkl)$ employed by the generation workflow. Inequivalent terminations were generated by shifting the cleavage 
plane along the surface normal. These cuts are labeled as $(hkl)$-$t_i$, where $t_i$ identifies the termination within the generated 
dataset. The initial pools contained 72 GaN, 81 LaN, and 73 ZnO surface models, giving 226 candidates and spanning 29 distinct orientations 
per compound. The surfaces were constructed with a thickness of approximately $20$~\AA~ and an initial vacuum separation of at least 
$20$~\AA. The resulting surface cells contained between 24 and 156 atoms (see Supplementary Table~1 for additional details). The sampled 
configurations comprised clean, stoichiometric cleavage terminations relaxed within their generated surface cells. Large-period 
reconstructions and the impact of crystalline defects and molecular adsorbates were not explored.

All surface candidates were automatically generated with the \verb!SurfAIcing! software \cite{surfaicing,Lopez2026} and geometrically 
relaxed with MACE \cite{batatia2022mace,batatia2025foundation}. A subset of 97 candidates, comprising 34 GaN, 30 LaN, and 33 ZnO models, 
was subsequently reoptimized with DFT. The basal polar $(001)$ terminations were excluded because their stability depends on surface-charge 
compensation, which can involve electronic charge redistribution, reconstruction, changes in stoichiometry, or the presence of molecular 
adsorbates \cite{Staemmler2003PolarZnO,Du2008PolarZnO}. The five lowest-energy candidates of each compound were retained for subsequent 
hybrid-functional DFT analysis of their electronic properties. Initial structural relaxation and energetic screening employed the medium 
MACE-MPA-0 foundation model \cite{batatia2025foundation}, interfaced with the atomic simulation environment (ASE) \cite{larsen2017ase}. 
A FIRE optimization was first applied to reduce large residual forces, followed by an L-BFGS optimization. Atomic positions were relaxed 
with fixed lattice vectors until the maximum force was below $0.004~\mathrm{eV\,\text{\AA}^{-1}}$. The resulting structures provided 
starting geometries for the subsequent DFT calculations.

In all cases, the surface formation energy was calculated as:
\begin{equation}
\gamma= \frac{E_{\mathrm{surf}}-N_{\mathrm{surf}}E_{\mathrm{bulk}}^{\mathrm{atom}}}{2A},
\label{eq:surface_energy}
\end{equation}
where $E_{\mathrm{surf}}$ is the total energy of the periodic surface model relaxed with the corresponding method, 
$E_{\mathrm{bulk}}^{\mathrm{atom}}$ is the bulk reference energy per atom calculated with that method, $N_{\mathrm{surf}}$ is the number 
of atoms in the surface model, and \textit{A} is the area of one exposed face. The factor of two accounts for the number of exposed faces
in the simulation cells. For models with inequivalent exposed terminations, the expression above provides their average surface energy. 
MACE energies were used to construct preliminary surface-stability rankings (Supplementary Figs.~1--3); however, the final identification 
of the energetically most favourable surfaces reported in this work is based entirely on semilocal DFT calculations.

\begin{figure*}[t]
\centering
\includegraphics[width=\textwidth]{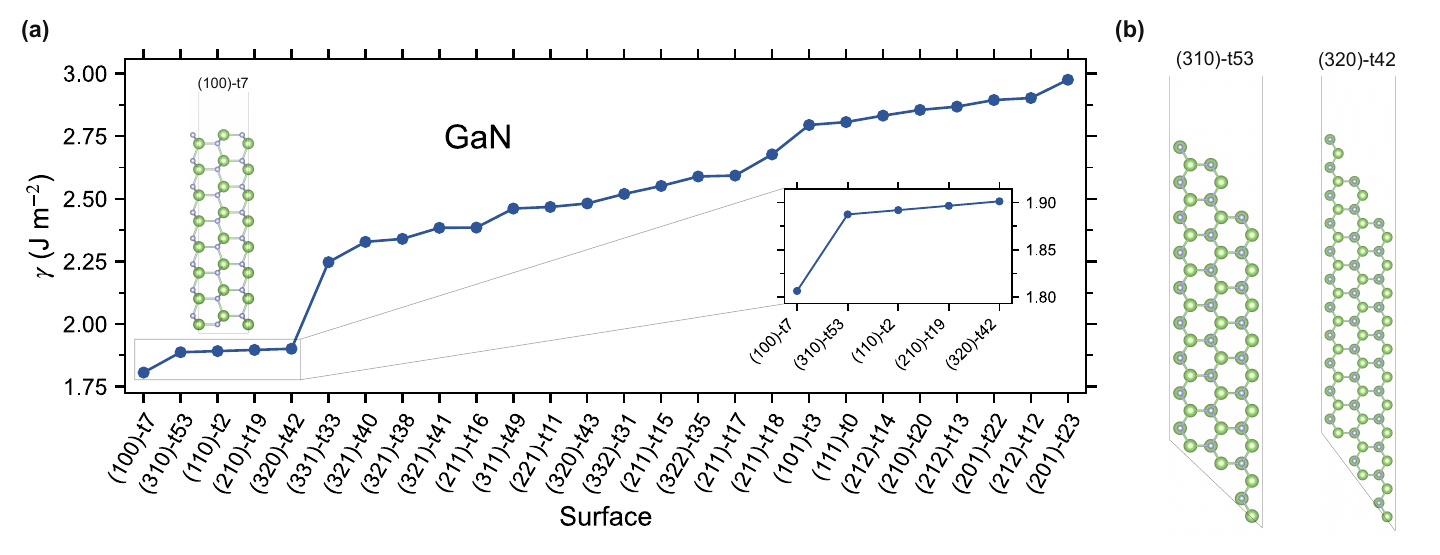}
\caption{(a)~DFT surface formation energies of representative wz-GaN surface models, ordered by increasing $\gamma$. The inset shows 
the five lowest-energy candidates. The atomic structure of the lowest-energy surface, $(100)$-t7, is also shown. (b)~Relaxed atomic 
structures of the high-index $(310)$-t53 and $(320)$-t42 surfaces. Green and light-blue spheres denote Ga and N atoms, respectively. 
Surface labels identify the Miller orientation and termination.}
\label{fig2}
\end{figure*}

\subsection{First-principles calculations}
\label{sec:dft}
First-principles DFT calculations were performed using the \verb!VASP! software \cite{kresse1996vasp,kresse1996efficient} using the 
projector-augmented-wave method \cite{blochl1994paw,kresse1999paw}. Exchange and correlation were described with the PBEsol functional 
\cite{perdew2008pbesol}, and the following electrons were considered in valence: Ga 4$s$ 4$p$; La 5$s$ 5$p$ 6$s$; Zn 4$s$ 3$d$; N 2$s$ 
2$p$; O 2$s$ 2$p$. Bulk structures were relaxed by optimizing both the unit cell and the atomic positions; the maximum force threshold 
was $0.004$~eV/\AA~. For surface models, atomic positions were relaxed while the lattice vectors were held fixed. Calculations were 
non-spin-polarized and used a plane-wave kinetic-energy cutoff of $600$~eV. Brillouin zone integrations employed $\Gamma$-centered regular 
meshes \cite{monkhorst1976special}, with one $k$-point along the third reciprocal-lattice direction of the surface cells. The meshes 
used for each material and orientation, together with those used for the bulk calculations, are reported in Supplementary Tables~2--3. 
Electronic self-consistency was converged to $10^{-7}$~eV using a Gaussian smearing of $0.1$~eV. Spin-orbit effects were systematically
disregarded. 

Dipole corrections were applied along the surface normal direction to correct artificial electrostatic interactions between periodically 
repeated surfaces \cite{neugebauer1992dipole}. The dipole center was defined separately for each structure from the center of its 
atom-containing region along the surface normal. These corrections address periodic-image effects while retaining physical surface dipoles 
and any internal field associated with the polarity and charge distribution of the surface model. Final PBEsol surface energies were 
evaluated using Eq.~\eqref{eq:surface_energy}, with consistently calculated surface and bulk reference energies. 

Electronic properties were calculated using the HSEsol screened hybrid functional \cite{Schimka2011,Krukau2006}, involving a short-range 
exact-exchange fraction of 25\% and a screening parameter of $0.2~\text{\AA}^{-1}$. HSEsol calculations were 
performed for the three bulk compounds and the five lowest-energy facets of each compound using their PBEsol-optimized geometries. 
The calculations were initialized from converged PBEsol electronic states and used a plane-wave cutoff of $520$~eV, Gaussian smearing 
of $0.1$~eV, and an electronic convergence threshold of $10^{-7}$~eV. Electronic density of states (DOS) and orbital-projected DOS 
(pDOS) were evaluated from the converged electronic states using the tetrahedron method with Bl\"ochl corrections. Vacuum-referenced 
band edges were obtained using a bulk-to-surface potential lineup, as done in previous works~\cite{Kahn2016,hinuma2019band,liu20,liu22}.

\begin{figure*}[t]
\centering
\includegraphics[width=\textwidth]{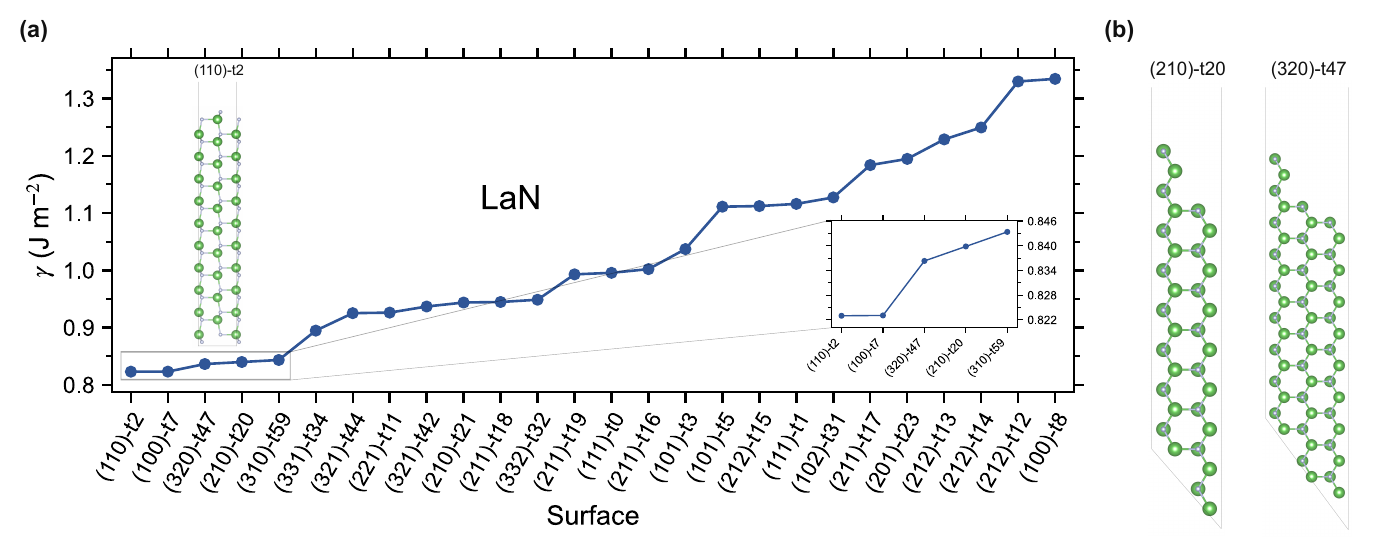}
\caption{
(a)~DFT surface energies of representative wz-LaN surface models, ordered by increasing $\gamma$. The inset shows the five lowest-energy 
candidates, including the nearly degenerate $(110)$-t2 and $(100)$-t7 surfaces. The atomic structure of $(110)$-t2 is also shown.
(b)~Relaxed atomic structures of the high-index $(210)$-t20 and $(320)$-t47 surfaces. Green and light-blue spheres denote La and N atoms, 
respectively. Surface labels identify the Miller orientation and termination.}
\label{fig3}
\end{figure*}

\section{Results}
\label{sec:results}
We investigated the surfaces of wurtzite GaN, LaN, and ZnO beyond the usual focus on high-symmetry low-index facets, sampling a large 
set of surfaces that includes low-symmetry high-index orientations (Methods). In the following sections, we present the surface-energy 
rankings and electronic properties obtained for each compound from first-principles DFT calculations, with particular emphasis on high-index 
facets and excluding polar surfaces like $(001)$. A comparison between the full surface-energy rankings obtained with DFT and 
with MACE is provided in Supplementary Figs.~1--3.

\begin{table*}[t]
\caption{Calculated bulk properties of wurtzite GaN, LaN, and ZnO as compared with experimental literature values. 
Equilibrium lattice parameters were obtained with the semilocal PBEsol functional and the band gap with the hybrid 
HSEsol functional (Methods). Reference band-gap values are experimental for GaN and ZnO and calculated with a DFT 
hybrid functional for LaN.}
\small
\setlength{\tabcolsep}{4pt}
\begin{ruledtabular}
\begin{tabular}{lccccccc}
Material
& $a^{\mathrm{PBEsol}}$
& $a^{\mathrm{Exp.}}$
& $c^{\mathrm{PBEsol}}$
& $c^{\mathrm{Exp.}}$
& $E_g^{\mathrm{HSEsol}}$
& $E_g^{\mathrm{Exp.}}$
& Band-gap character \\
& (\AA) & (\AA) & (\AA) & (\AA)
& (eV) & (eV) & \\
\hline
wz-GaN
& 3.18
& 3.19~\cite{vurgaftman2001band}
& 5.18
& 5.19~\cite{vurgaftman2001band}
& 3.20
& 3.40--3.50~\cite{vurgaftman2001band,monemar1974fundamental}
& direct \\

wz-LaN
& 4.12
& 4.08~\cite{krause2018situ}
& 5.80
& 5.84~\cite{krause2018situ}
& 2.10
& 2.29~\cite{rowberg2021structural}
& indirect \\

wz-ZnO
& 3.24
& 3.25~\cite{ozgur2005comprehensive}
& 5.22
& 5.20~\cite{ozgur2005comprehensive}
& 3.10
& 3.30~\cite{srikant1998optical}
& direct \\
\end{tabular}
\end{ruledtabular}
\label{tab1}
\end{table*}

\subsection{General properties}
\label{subsec:general}
Wurtzite crystallizes in the noncentrosymmetric hexagonal space group $P6_3mc$ (Fig.~\ref{fig1}a), with two formula units per primitive 
cell. Each cation is tetrahedrally coordinated by four anions, and vice versa. The structure can be viewed as two interpenetrating hexagonal 
close-packed sublattices, stacked in an $ABAB$ sequence along the polar $[0001]$ axis and displaced relative to each other along $c$. It is 
described by the lattice parameters $a$ and $c$ and their ratio $c/a$. The ideal tetrahedral arrangement corresponds to $c/a = \sqrt{8/3} 
\approx 1.633$. Deviations from this value reflect the departure from perfect tetrahedral bonding, and they modulate the spontaneous 
polarization and piezoelectric response discussed in the Introduction.

Table~1 compares the equilibrium lattice parameters obtained with PBEsol (Methods) and the experimental values. For GaN and ZnO the 
agreement is excellent. The calculated $a$ and $c$ parameters differ from experiment by 0.3\% and 0.2\% for GaN (3.18 and 5.18~\AA\ 
versus 3.19 and 5.19~\AA), and by 0.3\% and 0.4\% for ZnO (3.24 and 5.22~\AA\ versus 3.25 and 5.20~\AA). The corresponding $c/a$ ratios 
(1.629 for GaN and 1.611 for ZnO) are close to the ideal value and to the experimental ones. For LaN, the deviations are somewhat larger 
but still small. The $a$ parameter is overestimated by 1.0\% (4.12 versus 4.08~\AA) and $c$ is underestimated by 0.7\% (5.80 versus 
5.84~\AA). The resulting $c/a$ ratio of 1.41 differs markedly from that of GaN and ZnO and from the ideal value. This indicates that the
wurtzite phase of LaN is considerably distorted with respect to the ideal tetrahedral geometry, most likely owing to the large size of 
the La cation. Overall, PBEsol reproduces the structural parameters of the three compounds to within about 1\%, which supports its use 
for the relaxation of the surface models and for the surface-energy rankings.

Regarding the electronic structure, the HSEsol band gaps (Methods) are 3.20, 2.10, and 3.10~eV for GaN, LaN, and ZnO, respectively. GaN 
and ZnO have direct gaps, whereas LaN has an indirect one. For GaN and ZnO, the calculated values slightly underestimate the experimental 
gaps of 3.40--3.50~eV and 3.30~eV, by about 0.2--0.3~eV (6--9\%) and 0.2~eV (6\%), respectively. This residual discrepancy is small 
compared with the large errors typically obtained with semilocal functionals. It is also partly attributable to effects not included in 
our calculations, such as electron-phonon renormalization of the gap and the fact that the experimental values are optical gaps measured 
at finite temperature \cite{benitez2025a,benitez2025b,benitez2026a,benitez2026b}. For LaN, no experimental gap is available, so the 
reference in Table~1 is a previously reported hybrid-functional value of 2.29~eV. Our result lies 0.19~eV (about 8\%) below it. Overall, 
HSEsol provides a consistent and reliable description of the bulk band gaps of the three compounds, and the bulk references used in the 
band-alignment analysis below are therefore adequate.

\begin{figure*}[t]
\centering
\includegraphics[width=\textwidth]{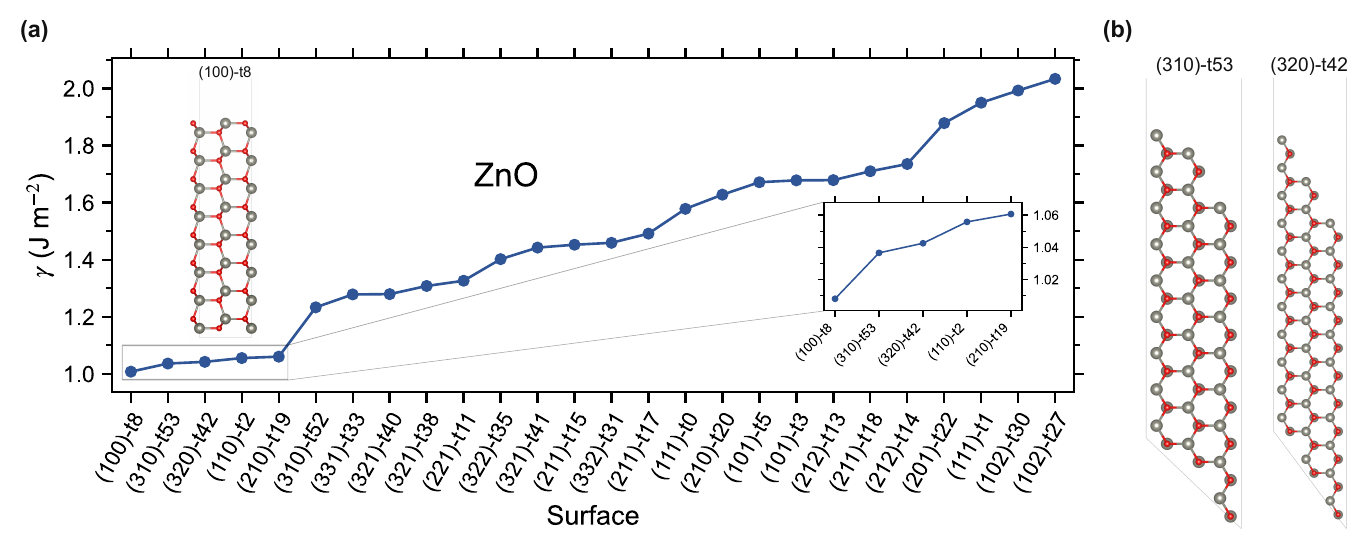}
\caption{
(a)~DFT surface energies of representative wz-ZnO surface models, ordered by increasing $\gamma$. The inset resolves the five lowest-energy
candidates. The atomic structure of the lowest-energy surface, $(100)$-t8, is also shown. (b)~Relaxed atomic structures of the high-index
$(310)$-t53 and $(320)$-t42 surfaces. Gray and red spheres denote Zn and O atoms, respectively. Surface labels identify the Miller
orientation and termination.}
\label{fig4}
\end{figure*}

\subsection{Surface formation energy}
\label{subsec:ranking}
For wz-GaN, the lowest calculated surface energy is obtained for $(100)$-t7, with $\gamma=1.806$~J\,m$^{-2}$ as shown in Fig.~\ref{fig2}a. 
It is followed by $(310)$-t53, $(110)$-t2, $(210)$-t19, and $(320)$-t42, with energies of $1.887$, $1.892$, $1.897$, and $1.901$~J\,m$^{-2}$,
respectively. These five candidates occupy a window of $0.095$~J\,m$^{-2}$, or approximately $5.3\%$ of the minimum value. In the three-index
notation used here, $(100)$ and $(110)$ correspond to the conventional $m$- and $a$-planes, respectively. Their calculated surface-energy 
ordering agrees with previous first-principles studies \cite{dreyer2014absolute,li2015computing}. Three members of the leading group are 
high-index prismatic surfaces, and $(310)$-t53 (Fig.~\ref{fig2}b) has a slightly lower calculated energy than the conventional $(110)$-t2 
termination. Thus, extending the sampled orientation space to low-symmetry high-index surfaces introduces several competing facets without 
a large energetic penalty. 

The low-energy region is narrower for wz-LaN as shown in Fig.~\ref{fig3}a. The conventional $(110)$-t2 and $(100)$-t7 surfaces have 
both calculated energies of $0.823$~J\,m$^{-2}$, hence are indistinguishable within our numerical resolution. The high-index 
$(320)$-t47, $(210)$-t20, and $(310)$-t59 candidates (Fig.~\ref{fig3}b) follow at $0.836$, $0.840$, and $0.843$~J\,m$^{-2}$. The five 
surfaces span only $0.020$~J\,m$^{-2}$, approximately $2.5\%$ of the minimum. These small separations identify a closely competing group 
of conventional and low-symmetry high-index terminations. In particular, the nearly identical energies of the two conventional surfaces 
provide little basis for favoring one orientation from surface energy alone. Previous first-principles studies of wz-LaN have established 
its bulk phase stability, electronic and polarization properties, point-defect physics, and heterostructure potential 
\cite{rowberg2021structural,rowberg2024point,rowberg2025heterostructure}. The present results extend this description to termination-resolved
surface energetics.

For wz-ZnO, $(100)$-t8 is the lowest-energy candidate, with $\gamma=1.008$~J\,m$^{-2}$, as shown in Fig.~\ref{fig4}a. The next four surfaces 
are $(310)$-t53, $(320)$-t42 (Fig.~\ref{fig4}b), $(110)$-t2, and $(210)$-t19, with energies of $1.037$, $1.043$, $1.056$, and 
$1.061$~J\,m$^{-2}$, respectively. Their energy window is $0.0525$~J\,m$^{-2}$, corresponding to $5.2\%$ of the minimum. Both $(310)$-t53 
and $(320)$-t42 lie below the conventional $(110)$-t2 surface in the calculated ranking. The calculated $m$- and $a$-plane surface energies 
agree closely with the values of 0.99 and 1.05~J\,m$^{-2}$, respectively, reported by Mora-Fonz \textit{et al} \cite{MoraFonz2015ZnO}. 
This agreement provides a benchmark for the calculations of higher-index surfaces.

Across the three wurtzite compounds, nine of the fifteen candidates selected for electronic analysis are low-symmetry high-index surfaces, 
and all fifteen are prismatic $(hk0)$ cuts. Their normals are perpendicular to the wurtzite $c$ axis, so the normal contribution from the 
bulk spontaneous polarization vanishes \cite{Stengel2011,Dreyer2016Polarization}. This geometric condition does not eliminate microscopic 
surface dipoles associated with local bonding and relaxation. The structures shown in Figs.~\ref{fig2}b--\ref{fig4}b contain both cations 
and anions in corrugated exposed arrangements. Their calculated energies demonstrate that such configurations can compete with conventional 
high-symmetry low-index terminations within the sampled, clean, stoichiometric surface models. This behavior is consistent with the 
capacity of wurtzite surfaces to accommodate undercoordination through structural relaxation, as established for conventional GaN and ZnO 
surfaces \cite{meyer2003density,Landmann2015GaN,MoraFonz2015ZnO}. Restricting the search to conventional high-symmetry low-index facets 
would omit several energetically competitive candidates in each compound.

\begin{figure*}[t]
\centering
\includegraphics[width=0.95\textwidth]{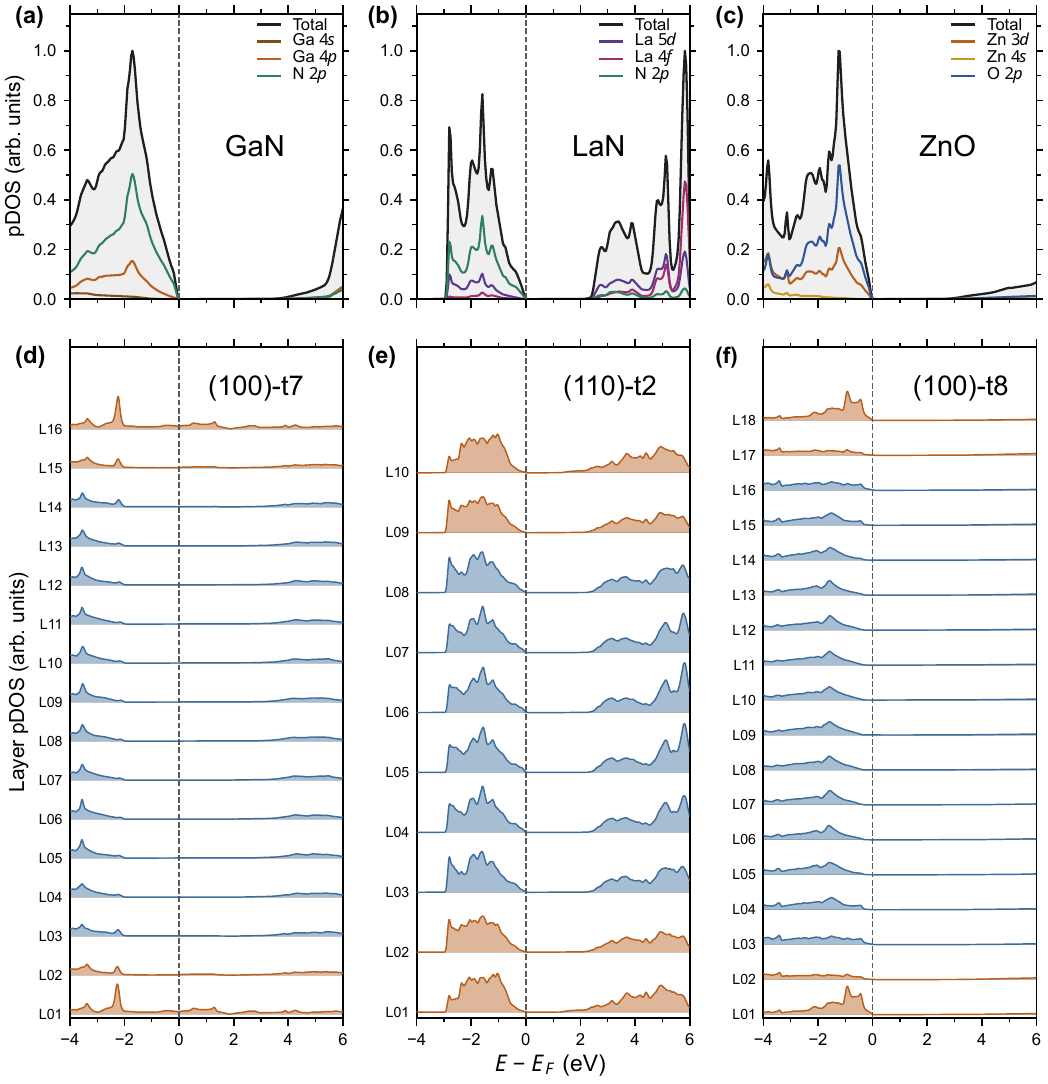}
\caption{HSEsol density of electronic states for bulk and surface models. (a)--(c) Total and orbital-projected densities of states
for bulk wz-GaN, wz-LaN, and wz-ZnO. (d)--(f) Layer-resolved projected densities of states for GaN $(100)$-t7, LaN $(110)$-t2, and 
ZnO $(100)$-t8, respectively. Orange curves highlight the outermost two atomic layers at each exposed face; blue curves denote the 
remaining layers. Layer spectra are normalized per atom, scaled by a common maximum within each surface model, and vertically offset 
for clarity. Labels L01, L02, and so forth identify successive atomic layers along the surface normal. All Fermi energy levels have 
been shifted to zero.}
\label{fig5}
\end{figure*}

\subsection{Electronic properties}
\label{subsec:electronic}
After establishing the structural and energy features for each wurtzite surface, we turn our attention to their electronic properties. 
Semilocal exchange-correlation functionals commonly underestimate semiconductor band gaps and may misplace localized \textit{d} and 
\textit{f} states due to self-interaction and delocalization errors 
\cite{perdew1985density,perdew1981self,mori2008localization,cohen2012challenges}. We therefore used the HSEsol screened hybrid functional 
for the electronic-structure calculations, which is designed for solids and is consistent with the PBEsol structural reference used 
throughout this work \cite{Schimka2011,Krukau2006} (Methods). 

The bulk orbital-projected densities of states in Figs.~\ref{fig5}a--c reveal the orbital character of the band-edge electronic states 
in the three compounds. In GaN, the valence-band edge is dominated by N $2p$ states, while Ga $4s$ and $4p$ states contribute to 
the conduction bands. The predominantly N $2p$ character near the valence-band edge agrees with the first-principles and spectroscopic 
analysis of wurtzite GaN by Magnuson \textit{et al} \cite{magnuson2010electronic}. In LaN, the valence-band edge is also mainly N $2p$ 
whereas the low-lying conduction states have predominantly La $5d$ character, with La $4f$ contributions appearing at higher energies. 
The N $2p$/La $5d$ band-edge character agrees with previous hybrid-functional calculations \cite{rowberg2021structural}. On the other hand,
ZnO exhibits an O $2p$-dominated valence-band edge and a predominantly Zn $4s$ conduction-band edge. The occupied Zn $3d$ states lie deeper 
in the valence bands and hybridize with O $2p$ states. This valence-band character is consistent with previous first-principles studies of 
ZnO \cite{uddin2006theoretical,karazhanov2006coulomb}.

The orbital-projected densities of states for the five lowest-energy surfaces of each compound are presented in Supplementary Figs.~4--6. 
The occupied states retain predominantly anion $2p$ character, while the spectral shapes and relative orbital contributions vary among the 
surface terminations. For GaN, the $(100)$-t7 surface exhibits a particularly distinct profile, including weak spectral weight near $E_F$ 
that is absent from the corresponding energy region of the other four candidates. The layer-resolved analysis below examines the spatial 
origin of these features for the lowest-energy surface of each compound.

Figures~\ref{fig5}d--f show the layer-resolved electronic contributions for the lowest-energy surface of each compound. For wz-GaN 
$(100)$-t7, the spectral weight near $E_{F}$ is concentrated in the outermost layers and strongly suppressed in the interior, indicating 
that these electronic states are localized at the exposed faces. This behavior is consistent with the modified bonding and relaxation of 
undercoordinated surface atoms. Previous first-principles calculations of the GaN $m$-plane identified occupied N-derived and unoccupied 
Ga-derived dangling-bond states associated with the relaxed surface structure \cite{Landmann2015GaN}.

The LaN $(110)$-t2 and ZnO $(100)$-t8 profiles also show changes in spectral shape and intensity between the exposed and interior layers. 
For LaN, occupied and unoccupied features extend through the interior, with modified intensities at the outer layers. For ZnO, the occupied 
edge develops distinct surface-layer features while the interior spectra remain similar to one another. These profiles characterize where 
the electronic contributions reside, a description that can be of practical interest for catalytic and microelectronic applications.

\begin{figure*}[t]
\centering
\includegraphics[width=\textwidth]{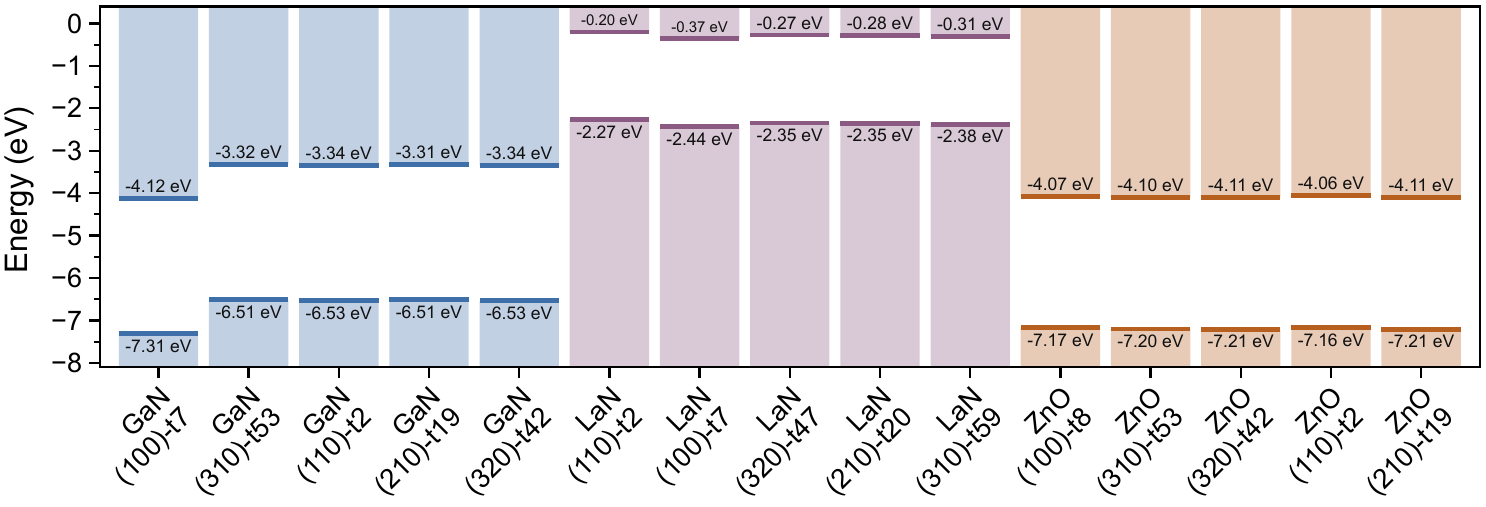}
\caption{Vacuum-referenced bulk band edges for the five lowest-energy surface models of wz-GaN (blue), wz-LaN (purple), and wz-ZnO (orange). 
Horizontal markers indicate the aligned valence-band maximum and conduction-band minimum. The vacuum level is set to zero. Band alignments 
were obtained from HSEsol calculations (Methods).}
\label{fig6}
\end{figure*}

\subsection{Band alignments}
\label{subsec:bandalign}
Figure~\ref{fig6} compares the vacuum-referenced band edges obtained by aligning the HSEsol bulk references through the local potential 
averaged in the bulk-like interior of each surface model (Methods). Within a given compound, the VBM and CBM shift together because their 
separation is fixed by the bulk gap used in the calculation. 

GaN exhibits the largest orientation dependence among the three compounds. The $(100)$-t7 orientation gives the deepest band edges, with 
$E_{\mathrm{VBM}}^{\mathrm{vac}}=-7.312$~eV and $E_{\mathrm{CBM}}^{\mathrm{vac}}=-4.117$~eV. The remaining four surfaces cluster between 
$-6.535$ and $-6.509$~eV for the VBM and between $-3.340$ and $-3.315$~eV for the CBM. The band-edge positions vary by $0.802$~eV across 
all five surfaces, while the four surfaces other than $(100)$-t7 differ by only $0.025$~eV. In particular, $(100)$-t7 and $(310)$-t53 
differ by $0.081$~J\,m$^{-2}$ in surface energy but by $0.799$~eV in their aligned band edges. The electron affinity of $4.12$~eV obtained 
for $(100)$-t7 is also in good agreement with the experimental $m$-plane value of $4.06 \pm 0.07$~eV \cite{Portz2018GaN}.

LaN shows a smaller, but appreciable, variation. The deepest alignment occurs for $(100)$-t7 with VBM and CBM values of $-2.438$ and 
$-0.366$~eV, respectively. The shallowest corresponds to $(110)$-t2, with values of $-2.275$ and $-0.203$~eV. Their band-edge difference 
is $0.163$~eV despite their nearly identical surface energies. The high-index $(320)$-t47, $(210)$-t20, and $(310)$-t59 surfaces lie 
between these limits. These results illustrate that almost equal surface energies can coexist with appreciably different vacuum-referenced 
band edges. 

By contrast, the five ZnO alignments are closely grouped. Their VBMs range from $-7.208$ to $-7.163$~eV, and their CBMs from $-4.108$ to 
$-4.063$~eV, giving a total variation of $0.045$~eV. The numerical extremes correspond to $(210)$-t19 and $(110)$-t2, respectively. For 
the lowest-energy $(100)$-t8 surface, the VBM and CBM are $-7.173$ and $-4.073$~eV. The $(100)$-t8 and $(310)$-t53 surfaces differ by 
$0.024$~eV in their band-edge positions and by 0.029~J\,m$^{-2}$ in surface energy.

\section{Discussion}
\label{sec:discussion}
The calculated surface energies show that low-symmetry high-index prismatic orientations can compete energetically with the conventional 
high-symmetry nonpolar faces of wurtzite GaN, LaN, and ZnO. Three of the five lowest-energy candidates of each compound are high-index 
surfaces, and these five candidates span only 2.5--5.3\% of the corresponding minimum surface energy. Restricting the search to the 
conventional $m$- and $a$-planes would therefore omit several competitive terminations. Determining whether these candidates appear as 
exposed equilibrium crystal facets requires an additional assessment of crystal morphology.

The competitiveness of these candidates also illustrates the limitations of using Miller indices alone to anticipate surface stability. 
The energetic cost of exposing a particular orientation depends on its atomic termination and the structural relaxation of undercoordinated 
atoms. Previous calculations of conventional GaN and ZnO surfaces have demonstrated substantial changes in local bonding upon relaxation
\cite{Landmann2015GaN,meyer2003density,MoraFonz2015ZnO}. The corrugated arrangements found among the present low-energy high-index candidates
are consistent with this capacity for structural adjustment. Their calculated energies establish their competitiveness within the sampled 
configurations, while the specific bonding changes responsible for stabilizing each termination remain to be resolved.

The MACE--DFT comparison (Supplementary Figs.~1--3) supports a workflow in which broad structural exploration performed with MLIP is 
followed by first-principles refinement. MACE provides preliminary energetic assessments and relaxed starting geometries, enabling a 
larger set of orientations and terminations to be examined before selecting candidates for DFT calculations. Differences in absolute 
surface energies and detailed ordering nevertheless remain, becoming particularly relevant when several candidates occupy a narrow energy 
window. The role of MACE in this workflow is therefore to facilitate exploration and pre-relaxation, while the PBEsol calculations 
establish the final energetic comparisons.

Similar surface energies do not necessarily imply similar vacuum-referenced band edges. Surface energy measures the excess energetic 
cost of exposing a surface, whereas electronic alignment depends on the potential difference between the material interior and vacuum. 
Although the fifteen selected surfaces are nonpolar prismatic cuts, local bonding and relaxation can generate microscopic surface dipoles 
that modify this difference. Such effects have been reported for the ionization potentials of nonpolar oxide surfaces \cite{hinuma2019band}. 
Surface dipoles provide a plausible interpretation of the calculated orientation dependence, although their structural and electronic 
contributions were not evaluated separately. GaN exhibits the clearest distinction between its lowest-energy surface and the remaining 
candidates; LaN combines nearly degenerate conventional surfaces with distinguishable alignments, whereas the selected ZnO alignments 
remain closely grouped. Energetic ranking alone therefore provides insufficient information to select a surface with a desired ionization
potential or electron affinity.

The layer-resolved pDOS complements the band-alignment analysis by identifying where electronic states reside within the surface models. 
For GaN $(100)$-t7, spectral weight near the calculated Fermi level is concentrated in the outermost layers and strongly suppressed in 
the interior. This observation is consistent with previous calculations identifying occupied N-derived and unoccupied Ga-derived 
dangling-bond states on the relaxed $m$-plane surface \cite{Landmann2015GaN}. Surface-localized states can influence carrier trapping 
and interfacial transfer, but their localization alone does not establish whether they favor carrier utilization or recombination. 
Consequently, similar aligned bulk band edges do not necessarily imply equivalent surface electronic behaviour or reactivity.

The vacuum-referenced band edges provide an exploratory comparison with water-redox energetics. At $pH = 0$ and $T = 298$~K, the 
conventional electron-energy references for $\mathrm{H}^{+}/\mathrm{H}_{2}$ and $\mathrm{O}_{2}/\mathrm{H}_{2}\mathrm{O}$ are 
approximately $-4.44$ and $-5.67$~eV, respectively \cite{Trasatti1986}. All five selected GaN and ZnO surfaces have aligned 
conduction-band edges above the hydrogen-reduction level and valence-band edges below the water-oxidation level. By contrast, 
the calculated LaN valence-band edges lie above the water-oxidation level, indicating insufficient oxidation driving force within 
this clean-surface screening picture. These comparisons do not establish photocatalytic activity: water adsorption and dissociation 
can modify semiconductor energy-level alignment \cite{Kharche2014Alignment,Kharche2017Erratum}, while operating performance also 
depends on surface coverage, carrier dynamics, and reaction kinetics.

For heterostructure design, our results identify orientations worth considering alongside conventional growth faces. Previous work 
on wurtzite LaN has established the importance of lattice matching, strain, and polarization discontinuities when selecting compatible 
materials and interface orientations \cite{rowberg2025heterostructure}. The present calculations contribute information about the 
energetic accessibility and vacuum-referenced electronic alignment of specific surface terminations. This information can guide the 
selection of structures for explicit interface calculations. Quantitative band offsets at a buried interface would additionally 
require accounting for its bonding, strain, and charge redistribution.

\section{Conclusions}
\label{sec:conclusions}
We have carried out a systematic first-principles exploration of the surface energetics and electronic properties of wurtzite GaN, 
LaN, and ZnO that extends well beyond conventional high-symmetry low-index facets. By combining automated surface generation, 
MACE-based screening, and PBEsol refinement of 97 selected candidates out of 226 clean, stoichiometric models, we find that the
lowest-energy nonpolar surfaces of the three compounds are all prismatic $(hk0)$ cuts, and that low-symmetry high-index terminations, 
such as $(310)$-t53 and $(320)$-t42, compete with the conventional $m$- and $a$-planes within 2.5--5.3\% of the minimum surface energy. 
Restricting the search to high-symmetry low-index facets would therefore omit several energetically accessible terminations in each 
compound.

The HSEsol band-edge alignments show that surface energy and electronic structure are complementary descriptors. Energetically similar 
surfaces can display very different vacuum-referenced band edges, as in GaN (almost 0.80~eV) and, to a lesser extent, LaN (0.16~eV), 
while the ZnO surfaces remain closely grouped (0.045~eV). Likewise, layer-resolved densities of states reveal surface-localized states 
in the outermost layers of the GaN $(100)$-t7 surface. Our results thus call for assessing stability and band alignment jointly when
selecting surfaces for catalysis, sensing, or heterostructure growth. They also define a set of candidate surfaces for future work on 
reconstruction, adsorption, equilibrium crystal morphology, and interface formation, which lie beyond the clean, stoichiometric models 
considered here.

\section*{Acknowledgments}
C.C. acknowledges support by MICIN/AEI/10.13039/501100011033 and ERDF/EU under the grants CNS2025-165467, PID2023-146623NB-I00 and 
PID2023-147469NB-C21 and by the Generalitat de Catalunya under the grants 2021SGR-00343, 2021SGR-01519 and 2021SGR-01411. Computational 
support was provided by the Red Española de Supercomputación under the grants FI-2024-1-0005, FI-2024-2-0003, FI-2024-3-0004, 
FI-2024-1-0025, FI-2024-2-0006, and FI-2025-1-0015. J.Ll. is a Serra Húnter Fellow and gratefully acknowledges support from the Academia 
Excellence program and 2025 SGR 01006 from the Generalitat de Catalunya, and financial support from MICIU/AEI/10.13039/501100011033 and 
FEDER through project PID2024-156765OB-C21. This work is part of the Maria de Maeztu Units of Excellence Programme CEX2023-001300-M 
funded by MCIN/AEI (10.13039/501100011033).

\bibliography{bibliography}

\end{document}